# A comparison of Monte Carlo dropout and bootstrap aggregation on the performance and uncertainty estimation in radiation therapy dose prediction with deep learning neural networks

Dan Nguyen, Azar Sadeghnejad Barkousaraie, Gyanendra Bohara, Anjali Balagopal, Rafe McBeth, Mu-Han Lin, Steve Jiang

Medical Artificial Intelligence and Automation (MAIA) Laboratory, Department of Radiation Oncology, UT Southwestern Medical Center

Dan.Nguyen@UTSouthwestern.edu

Recently, artificial intelligence technologies and algorithms have become a major focus for advancements in treatment planning for radiation therapy. As these are starting to become incorporated into the clinical workflow, a major concern from clinicians is not whether the model is accurate, but whether the model can express to a human operator when it does not know if its answer is correct. We propose to use Monte Carlo dropout (MCDO) and the bootstrap aggregation (bagging) technique on deep learning models to produce uncertainty estimations for radiation therapy dose prediction. We show that both models are capable of generating a reasonable uncertainty map, and, with our proposed scaling technique, creating interpretable uncertainties and bounds on the prediction and any relevant metrics. Performance-wise, bagging provides statistically significant reduced loss value and errors in most of the metrics investigated in this study. The addition of bagging was able to further reduce errors by another 0.34% for $D_{mean}$ and 0.19% for $D_{max}$, on average, when compared to the baseline model. Overall, the bagging framework provided significantly lower MAE of 2.62, as opposed to the baseline model's MAE of 2.87. The usefulness of bagging, from solely a performance standpoint, does highly depend on the problem and the acceptable predictive error, and its high upfront computational cost during training should be factored in to deciding whether it is advantageous to use it. In terms of deployment with uncertainty estimations turned on, both methods offer the same performance time of about 12 seconds. As an ensemble-based metaheuristic, bagging can be used with existing machine learning architectures to improve stability and performance, and MCDO can be applied to any deep learning models that have dropout as part of their architecture.

# I. Introduction

The field of radiation therapy is ever-changing, with technologies and distributions of patient populations evolving over time. Some of the largest advancements in the past few decades have come from the creation and adoption of intensity modulated radiation therapy (IMRT) (Brahme, 1988; Bortfeld *et al.*, 1990; Bortfeld *et al.*, 1994; Webb, 1989; Convery and Rosenbloom, 1992; Xia and Verhey, 1998; Keller-Reichenbecher *et al.*, 1999) and volume modulated arc therapy (VMAT)(Yu, 1995; Otto, 2008; Palma *et al.*, 2008; Shaffer *et al.*, 2009; Shaffer *et al.*, 2010; Xing, 2003; Earl *et al.*, 2003; Daliang Cao and Muhammad, 2009), greatly improving treatment plan quality. In recent years, a major wave of innovations has been coming from the development of artificial intelligence (AI) technologies, and their application into radiation therapy. One particular area has been focused on the prediction of clinically relevant dosimetric endpoints, which can include single dose constraints, dose volume histograms (DVH), or the full volumetric dose distribution. The earlier adoption of machine learning technologies led to knowledge-based planning (KBP)(Zhu *et al.*, 2011; Appenzoller *et al.*, 2012; Wu *et al.*, 2014; Shiraishi *et al.*, 2015; Moore *et al.*, 2011; Shiraishi and Moore, 2016; Wu *et al.*, 2009; Wu *et al.*, 2011; Wu *et al.*, 2013; Tran *et al.*, 2017; Yuan *et al.*, 2012; Lian *et al.*, 2013; Folkerts *et al.*, 2016; Good *et al.*, 2013; Valdes *et al.*, 2017; Yang *et al.*, 2013), which focused on the use of historical patient data in order to predict single dose constraints or dose volume histograms. KBP required the careful selection of human engineered features to be calculated from the data in order to successfully predict the DVH and dose constraints.

Recently many deep learning (DL) dose prediction models have risen in literature that cover a range of sites and modalities including head-and-neck (H&N) IMRT(Fan *et al.*, 2019; Babier *et al.*, 2018b; Mahmood *et al.*, 2018; Babier *et al.*, 2018a), H&N VMAT(Nguyen *et al.*, 2019b), prostate IMRT(Kearney *et al.*, 2018; Nguyen *et al.*, 2019a; Nguyen *et al.*, 2017; Nguyen *et al.*, 2019d; Bohara *et al.*, 2020), prostate VMAT(Shiraishi and Moore, 2016), and lung IMRT(Barragán-Montero *et al.*, 2019). These methods have made great strides in the ability to directly predict the volumetric dose distribution. With the 3D dose distribution, it is possible to fully reconstruct the DVH and dose constraints. In addition, these DL-based methods were able to operate on the patient computed tomography (CT) image and contours directly, or other simple features, in order to learn and extract their own complex features for making an accurate prediction.

However, as we move closer to clinical implementation of the DL-based methods, concerns regarding safety in using the AI model on patients has risen greatly. One of the largest worries from physicians is not whether the model is accurate, but whether the model can express to a human operator when it does not know if its answer is correct. The ability for a model to output its uncertainty has been developed through Bayesian approaches(Beck and Katafygiotis, 1998), such as the Gaussian process(Rasmussen and Williams, 2006). However, these have been largely limiting due to the extremely high computational cost to implement. Recently, Gal and Ghahramani proposed that one can use a method called Monte Carlo Dropout (MCDO) during the evaluation phase to produce an uncertainty estimation(Gal and Ghahramani, 2016). They showed that this method approximates the Gaussian process by effectively sampling the model's weights from a Bernoulli distribution. The computational cost of implementing this method only comes in the evaluation phase, when the model needs to be evaluated multiple times to create Monte Carlo estimations. There is no additional cost during the training phase of the deep learning model.

Upon inspection of the MCDO technique, we realize another method called bootstrap aggregation (bagging)(Breiman, 1996), also generates multiple models that contain a distribution of weights, and therefore, can natively generate its own uncertainty estimate as well. Bagging is an ensemble-based machine learning technique that was originally designed for improving the stability and performance of deep learning models. The basic idea behind how it works is that many models are trained, where each model trains on a portion of the total training set. The training subset may be selected with or without replacement from the total training set. For evaluation, all models make a prediction and these predictions are equally averaged to generate the final result. Bagging has been shown to create a more stable model by reducing the chance of overfitting and the prediction variance. However, it is not too commonly used in typical deep learning scenarios due to most theoretical deep learning applications having a very large data pool with hundreds of thousands to millions of data points. However, bagging can be efficiently implemented in smaller dataset scenarios, such as for medical applications.

In this study, we will compare the performance of the baseline model and bagging framework, as well as their respective uncertainty estimations, and discuss the pros and cons of each method. Due to the relative nature of these uncertainties, we introduce a method to scale the raw uncertainties into interpretable values. We show how this scaled uncertainty can be used to provide tangible upper and lower bound estimates of the error.

## II. Methods
### II.A. Data

In this study we used a public H&N cancer patient data available from the OpenKBP – 2020 AAPM Grand Challenge(Babier *et al.*, 2020a; Babier *et al.*, 2020b). The dataset consisted of 200 training, 40 validation, and 100 testing H&N cancer patients. For each patient, the data available were the patient CT, masks of the planning target volumes (PTV) and organs-at-risk (OAR), the 3D dose distribution, and the voxel spacing information. The dose distribution was generated from an 9 beam equidistant coplanar IMRT setup with 6 MV fields. The structures available in the dataset were the PTV70, PTV63, PTV56, larynx, esophagus, mandible, brainstem, spinal cord, right parotid, and left parotid. Most patients did not have the complete set of structures, and the mode number of structures was 8 per patient. The specific details are broken down in Table 1.

| Structure | Number of Structures | | |
|---|---|---|---|
| | Training | Validation | Testing |
| **PTV70** | 200 | 40 | 100 |
| **PTV63** | 91 | 26 | 42 |
| **PTV56** | 178 | 38 | 91 |
| **Larynx** | 100 | 22 | 53 |
| **Esophagus** | 81 | 17 | 43 |
| **Mandible** | 136 | 28 | 72 |
| **Brainstem** | 182 | 35 | 89 |
| Spinal Cord | 188 | 38 | 88 |
| Right Parotid | 192 | 40 | 99 |
| Left Parotid | 193 | 38 | 98 |

Table 1: Exact breakdown of the number of available structures in the training, validation and testing set, which consisted of 200, 40, and 100 patients, respectively.

In addition, a "possible dose mask" was included in the dataset to indicate voxels where there can be non-zero dose. The CT, dose, and structures are all formatted as 128 x 128 x 128 arrays. The voxel spacing was not the same across different patients, but the mode voxel spacing was 3.906 mm x 3.906 mm x 2.5 mm, with a range (min, max) of (0.45, 4.18) mm x (0.45, 4.18) mm x (0.42, 2.58) mm.

Using the Boolean structure masks and the voxel spacing information, we generated a signed distance array for each structure. Each voxel from the signed distance array represents the physical distance from the closest edge of the contour. Voxels outside the contour had a positive distance value, voxels inside had a negative distance value.

## II.B. Deep Learning Framework

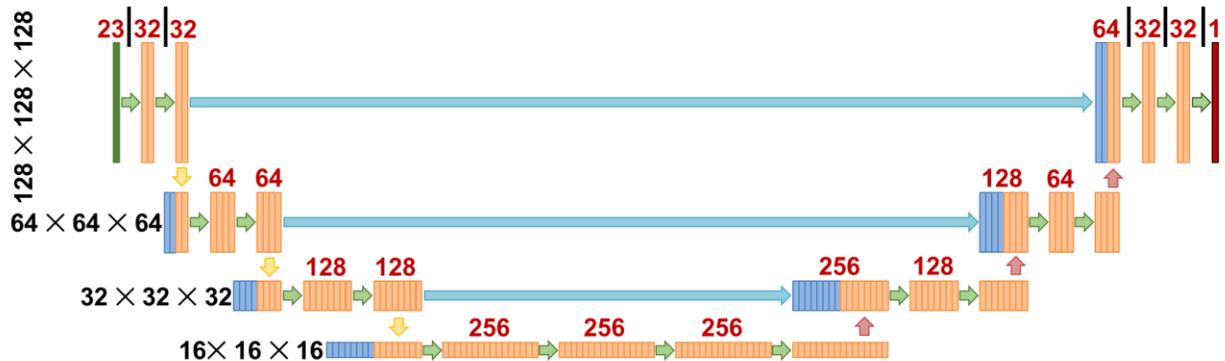

Figure 1: Schematic of the U-net used in the study. Black numbers to the left represent the size of each feature map at each level of the U-net. Red numbers above represent the number of feature maps that were calculated for that layer.

To keep the study as fair as possible, all models trained in this study utilized the same exact architecture and hyperparameters. In this study, we used a U-net style architecture(Ronneberger *et al.*, 2015). The U-net takes a 23 channel input of size 128 x 128 x 128 voxels. The model starts by performing a padded convolution with a kernel size of 3 x 3 x 3, followed by ReLU, Group Norm (GN)(Wu and He, 2018), and then DropBlock (DB)(Ghiasi *et al.*, 2018). In this paper, we will refer to this set of operations as Conv-ReLU-GN-DB. This set of operations was performed twice, and then two downsampling operations were applied—1) Max pooling and 2) a strided convolution with a kernel size of 2 x 2 x 2 and stride size of 2 x 2 x 2, ReLU, GN, and DB—and concatenated together. These sets of calculations were performed a total of 3 times to reach the bottom of the U-net, at which point 4 sets of Conv-ReLU-GN-DB were applied. Then the features were upsampled, which consists of a 2 x 2 x 2 upscale and one set of Conv-ReLU-GN-DB operations. This set of upsampling operations, followed by 2 x Conv-ReLU-GN-DB, were performed 3 times to bring the data back to its original resolution as the input. At each upsampling step, the features from the left side of the U-net were copied over and concatenated with the upsampled features. For DropBlock, the block size was set to 5 x 5 x 5, and the dropout rate was based on the level of the U-net: 0.119 (1st level (top)), 0.141 (2nd level), 0.168 (3rd level), 0.2 (4th level (bottom)). This was based on a heuristic formula $droprate = 0.2 * \left(\frac{current\ number\ of\ filters}{\max number\ of\ filters}\right)^{1/4}$, which was used in previous dose prediction study(Nguyen *et al.*, 2019c). The schematic of the U-net used in this study is shown in Figure 1.

## II.C. Training

Both the baseline model and bagging framework used the same training, validation, and testing data split as was used originally in the KBP competition. The baseline model utilized 200 patients for training and 40 for validation throughout its training process. For the bagging framework, a total of 24 models were trained from scratch. For each bagging model, we randomly selected 160 of the 200 training patients for training the model, and the remaining 40 were used for the observation of a validation loss for each model. The original 40 validation patients, that are the same as that of the baseline model, were used later for scaling the raw uncertainty, as described in section II.E. As input, the patient's CT, Boolean structure masks, and signed distance arrays for the 10 structures, and the possible dose mask were included. In total, the data formed a 23 channel input, and were used by the model to predict the 3D dose distribution.

Since the KBP competition originally based their evaluation metric on mean absolute error (MAE), we used a surrogate loss known as the Huber loss (Huber, 1992) defined as

$$H_\delta(x) = \begin{cases} \frac{1}{2}x^2 & for\ |x| \leq \delta \\ \delta\left(|x| - \frac{1}{2}\delta\right) & otherwise \end{cases}$$

(1)

From an optimization perspective, the Huber loss has similar characteristics to MAE, but is additionally differentiable at $x = 0$. For optimization, we used the Adam algorithm (Kingma and Ba, 2014) with learning rate of 0.001. The model is trained for a total of 100,000 iterations, with evaluation of the validation loss every 100 iterations. To help mitigate overfitting to the training data, the final model we take from each training session is the one that best minimizes the validation loss. The batch size was set to 1, due to memory limitations. The models were trained on a V100 GPU with 32 GB of memory, using TensorFlow version 2.1.

## II.D. Performance Evaluation

To evaluate the performance of the baseline model versus the bagging framework, we assessed its prediction accuracy on the 100 test patients. We viewed the voxel-wise MAE, and evaluated the effects of number of models used for bagging on the MAE. We additionally viewed the effects of the models on more clinically relevant metrics, which included the mean dose ($D_{mean}$) and max dose ($D_{max}$) to each structure. $D_{max}$ is defined as the dose to 2% of the volume for each structure, as recommended by the ICRU-83 report(Grégoire and Mackie, 2011). We also evaluated the model performance on the total PTV dose coverage $D_{99}$, $D_{98}$, $D_{95}$, which is the dose to 99%, 98%, and 95% of the total PTV volume. We assessed the PTV homogeneity $\left(\frac{D2-D98}{D50}\right)$ for each PTV and for the total PTV, as well the plan conformity $\left(\frac{(V_{PTV} \cap V_{100\%Iso})^2}{V_{PTV} \times V_{100\%Iso}}\right)$(Van't Riet *et al.*, 1997; Paddick, 2000) using the total PTV. We also investigated the isodose similarity, which is a Dice coefficient(Dice, 1945; Sorensen, 1948) calculation to compare the similarity of the isodose volumes. This was done by first selecting an isodose level and segmenting both the ground truth and predicted dose such that all the voxels within the volume contain that dose value and higher. Then the Dice similarity was calculated for these two volumes. This was repeated for all isodose levels from 0% to 100% of the prescription dose. In our case, we calculated the isodose similarity from 0 Gy to 70 Gy, at 0.1 Gy increments.

Pairwise t-tests were performed to compare the errors of the baseline model's prediction versus the errors of the bagging framework's predictions using the 100 test patients. The errors were calculated by finding the MAE between the prediction and the ground truth doses on several metrics, including the individual structures $D_{mean}$ and $D_{max}$, PTV dose coverage ($D_{99}$, $D_{98}$, $D_{95}$), PTV homogeneity, plan conformity, and the overall isodose similarity.

## II.E. Uncertainty Estimation

In order to obtain uncertainty estimation, the models produce a distribution of estimations and the distribution's standard deviation is measured as the uncertainty. In the MCDO method, the models were first trained with dropout turned on (Srivastava *et al.*, 2014). Dropout is a technique originally invented as a method for preventing overfitting. It works by, during each update iteration, stochastically turning off each model weight in the network with some probability, called the dropout rate. A higher dropout rate can better prevent the neural network from overfitting, with the potential cost of degrading the model accuracy. Historically, dropout was only applied during the training phase, and then turned completely off during the evaluation and deployment phase. For the MCDO method, the model generates multiple predictions while dropout is turned on. This randomly turns off a different portion of the network each iteration, causing the predictions to have some variation. A standard deviation can then be calculated and interpreted as the uncertainty.

For the bagging framework, multiple separate models are trained, each with a randomly selected 160 patients out of the 200 training patients. This difference, alongside other factors such as random weight initialization and random patient selection for each training iteration, affect the final convergence of each model's weights, and therefore, its predictions are slightly varied among the different trained models. The standard deviation among all of the bagging models' predictions can then be calculated.

In summary, the idea behind uncertainty estimation is that both the baseline model, through using MCDO, and the bagging framework are capable of producing a distribution of predictions. The uncertainty itself is the voxel-wise standard deviation of these predictions. Mathematically, for either method, the model's raw uncertainty (standard deviation), can be estimated by the equation:

$$uncertainty_{raw}(y^*, x, \{W_1, \cdots, W_T\}) \approx \sqrt{\frac{1}{T}\sum_{t=1}^{T}\hat{y}^*(x, W_t)^2 - \left(\frac{1}{T}\sum_{t=1}^{T}\hat{y}^*(x, W_t)\right)^2}$$

(2)

where $\hat{y}^*(x, W_t)$ is the trained model's prediction given an input $x$ and a $t$ set of weights, $W_t$. For the baseline method, by utilizing Dropout during inference, the model's weights, $W_t$, are sampled from a Bernoulli distribution via variational inference. For the bagging method, $W_t$, is taken from each trained model instance. As a standard deviation of the prediction, the units of raw uncertainty match the units of the prediction, which is Gy.

Due to the nature of MCDO for the baseline model, as well as the nature of the number of samples selected during training for bagging, the raw uncertainty realistically can only provide a relative measure since its magnitude is dependent on the aforementioned factors. Here, we propose to scale the $uncertainty_{raw}$ into a more interpretable scale. Specifically, we used the following calculation on the validation data to find a scaling factor $m_\sigma$ for a given region of interest (ROI) with $n_{ROI}$ number of voxels:

$$m_i = \frac{dose_{pred,i} - dose_{true,i}}{uncertainty_{raw,i}}$$

(3)

$$m_{\sigma,ROI} = \sqrt{\frac{1}{n_{ROI}}\sum_{i \in ROI} m_i^2 - \left(\frac{1}{n_{ROI}}\sum_{i \in ROI} m_i\right)^2}$$

(4)

This value, $m_\sigma$, can be multiplied with the $uncertainty_{raw}$ to generate $uncertaity_{scaled}$. This method to calculate the scaling works when the error distribution is approximately symmetric about zero, which is often the case for many deep learning algorithms that have learned using symmetric loss functions. We found this value separately for the PTVs and the rest of the body for each the baseline and bagging methods. In total we calculated 4 $m_\sigma$ values— $m_{\sigma,(Body-PTV),Baseline}$, $m_{\sigma,PTV,Baseline}$, $m_{\sigma,(Body-PTV),Bagging}$, $m_{\sigma,PTV,Bagging}$—that are used to scale the raw uncertainties in the PTVs and the body – PTV separately. We assessed the use of these uncertainties for bound estimations on the test patient data. In addition, we calculated the Pearson correlation coefficient between the estimated uncertainty and the prediction error.

# III. Results

## III.A. Performance Evaluation

Each model took an average of 57.2 hours to train on 1 GPU. For training the bagging framework of 24 models, this equates to 1372.8 hours. While there is a high upfront cost for training the bagging framework, the evaluation time is very short, with each prediction taking an average of 0.48 seconds per GPU. The 24 model bagging framework took less than 12 seconds to complete a prediction for 1 patient. As a single model, the baseline model took 57.2 hours to train, and an average of 0.48 seconds per prediction for evaluation.

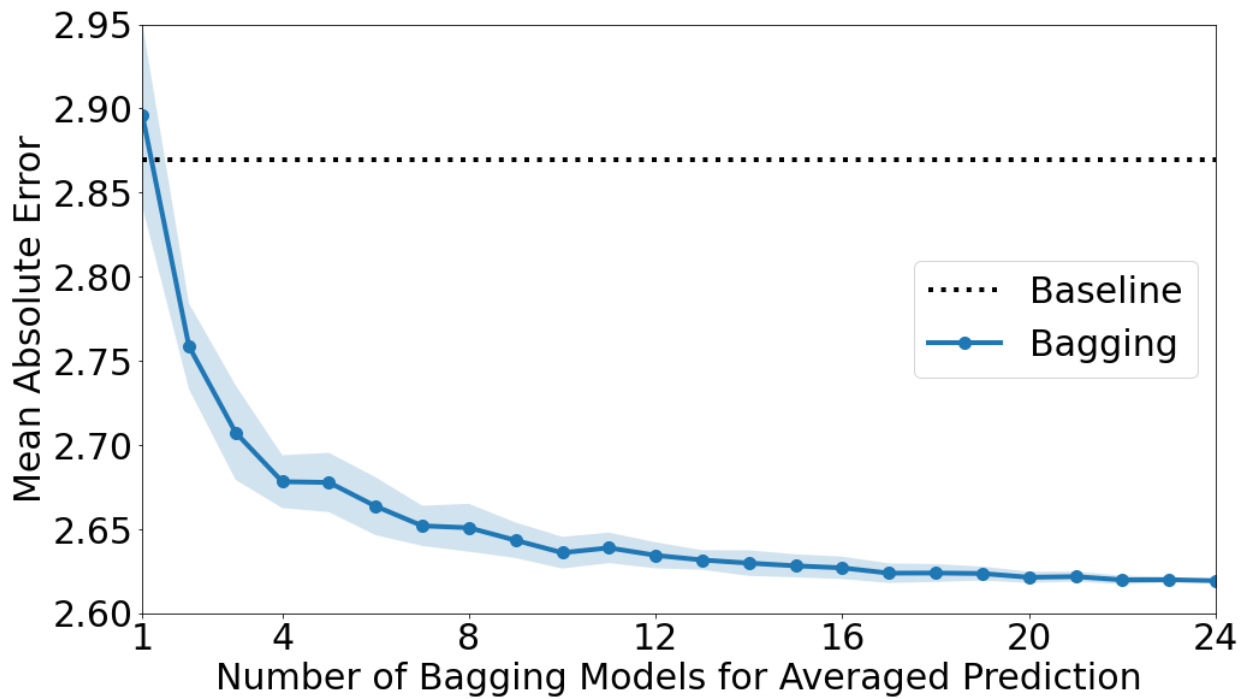

Figure 2: Evaluation of the effects of the number of models used in the bagging procedure on the Mean Absolute Error (MAE) for the 100 test patients. With the exception of the 24 model-averaged prediction, all other data points randomly chose *n* number of models to average the prediction, and this was repeated 24 times with unique combination of models. The error band represents the standard deviation.

The benefit of the bagging framework in MAE is illustrated in Figure 2. We can see that the baseline model performed slightly better than the mean performance of a single model from the bagging framework, likely due to the fact that the baseline model trained directly on 200 patients while each individual bagging model trained on 160 patients. There was substantial improvement in the MAE as multiple model's predictions are averaged together in the bagging framework, with a significant improvement with as little as 2 models. The improvement in MAE continued as more models added into the bagging framework, with diminishing returns. Overall, the baseline model and bagging framework had an average MAE of 2.87 and 2.62, respectively.

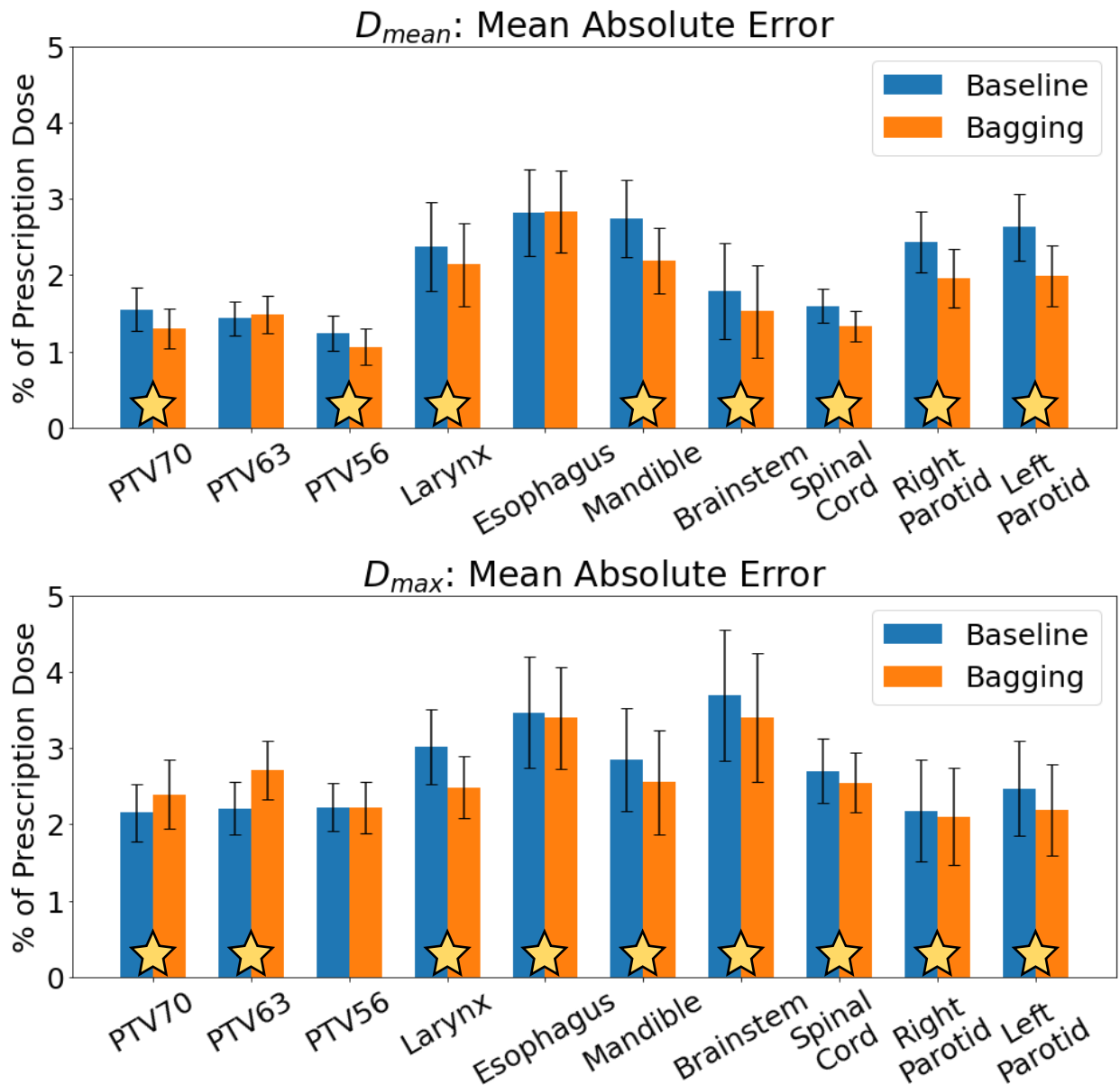

**Figure 3: Mean absolute error of the $D_{mean}$ (top) and $D_{max}$ (bottom) between the predictions and the ground truth dose. Error bar represents the 95% confidence interval $\left(\bar{x} \pm 1.96\frac{\sigma}{\sqrt{n}}\right)$. Statistically significant differences in the MAE are marked with a yellow star.**

Figure 3 shows the MAE of $D_{mean}$ and $D_{max}$, respectively. Overall, both models were capable of maintaining a mean absolute error of less than 3% and 4% of the prescription dose for $D_{mean}$ and $D_{max}$, respectively. Using a pairwise t-test on the 100 test patients, the MAE of $D_{mean}$ was significantly reduced (p < 0.05) for all of the structures except for PTV63 and esophagus, which had no significant difference. For $D_{max}$, except for the PTV56, all other differences were found to be statistically significant (p < 0.05). On average, the bagging model was able to reduce the MAE to the OARs by 0.24 Gy ($D_{mean}$) and 0.13 Gy ($D_{max}$), when compared to the baseline model.

The bagging framework predicted an average $D_{max}$ of 72.5 Gy, 70.6 Gy, and 66.8 Gy for the PTV70, PTV63, and PTV56 respectively, whereas the baseline model predicted 73.0 Gy, 71.1 Gy, and 66.9 Gy, respectively. The Ground Truth average $D_{max}$ was 73.7 Gy, 71.9 Gy, and 68.0 Gy, respectively. The larger error in $D_{max}$ to the PTVs is due to the averaging effect from multiple predictions on the max dose point, leading to a lower hotspot and a more homogeneous PTV.

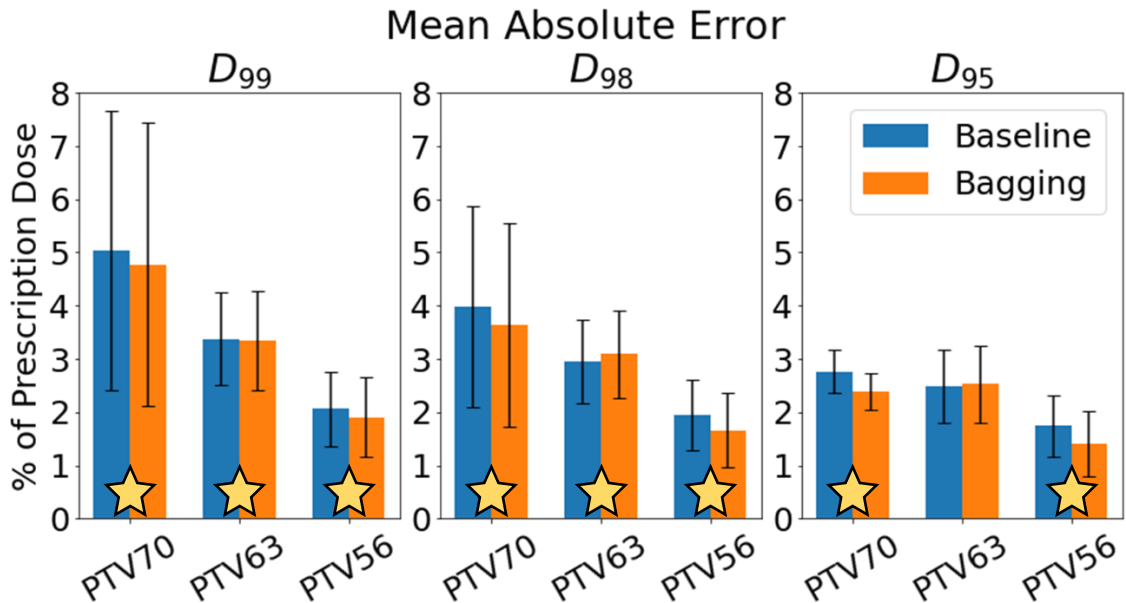

**Figure 4: Mean absolute error of the dose coverage ($D_{99}, D_{98}, D_{95}$) between the predictions and the ground truth dose. Error bar represents the 95% confidence interval $\left(\bar{x} \pm 1.96 \frac{\sigma}{\sqrt{n}}\right)$. Statistically significant differences in the MAE are marked with a yellow star.**

Figure 4 shows the MAE of the dose coverage. Overall, the baseline model and bagging frameworks performed similarly, with the largest average difference less than 0.4% between the baseline MAE and bagging MAE ($D_{95}$ of PTV56). All differences, except for $D_{95}$ of the PTV63, were found to be statistically significant (p <0.05). The only metric where bagging was statistically worse than baseline was the $D_{98}$ for the PTV63. However, the mean difference for this metric is 0.1% (0.07 Gy) of the prescription dose, which is clinically insignificant. On average, bagging had the largest benefit in reducing the dose coverage error to the PTV70 by an additional 0.26 Gy ($D_{99}$), 0.24 Gy ($D_{98}$), and 0.18 Gy ($D_{95}$).

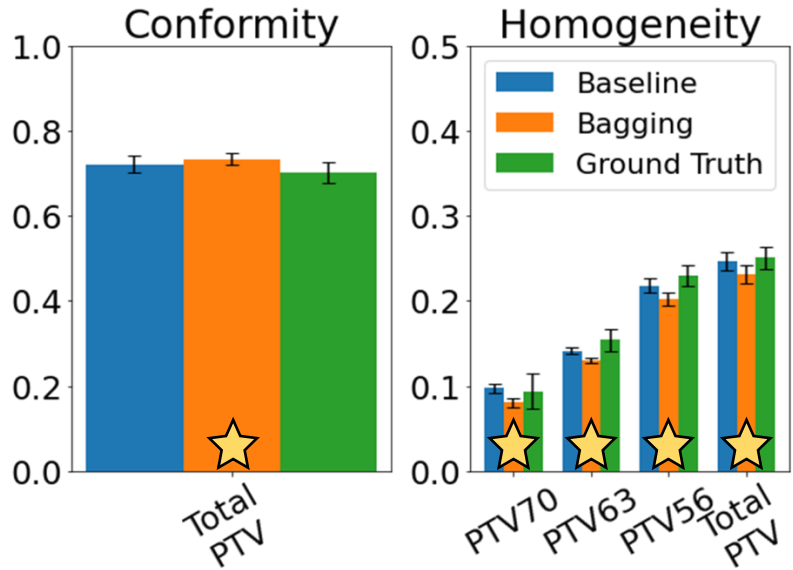

**Figure 5: Conformity and homogeneity of the baseline prediction, bagging prediction, and ground truth. Error bar represents the 95% confidence interval $\left(\bar{x} \pm 1.96\frac{\sigma}{\sqrt{n}}\right)$. All comparisons of the baseline prediction errors and bagging prediction errors were found to be statistically significant differences, which are marked with a yellow star.**

Figure 5 shows the conformity and homogeneity. Ideally, a perfect conformity is 1 and a perfect homogeneity is 0. All differences between the baseline prediction errors and bagging prediction errors were found to be statistically significant, with the bagging framework having larger errors than the baseline model. From a clinical standpoint, we found that the bagging framework is predicting a better conformity and homogeneity than both the ground truth and baseline prediction. The better homogeneity is likely caused by the averaging done from all 24 model's predictions in bagging, which washes out both hot and cold spots in the PTVs.

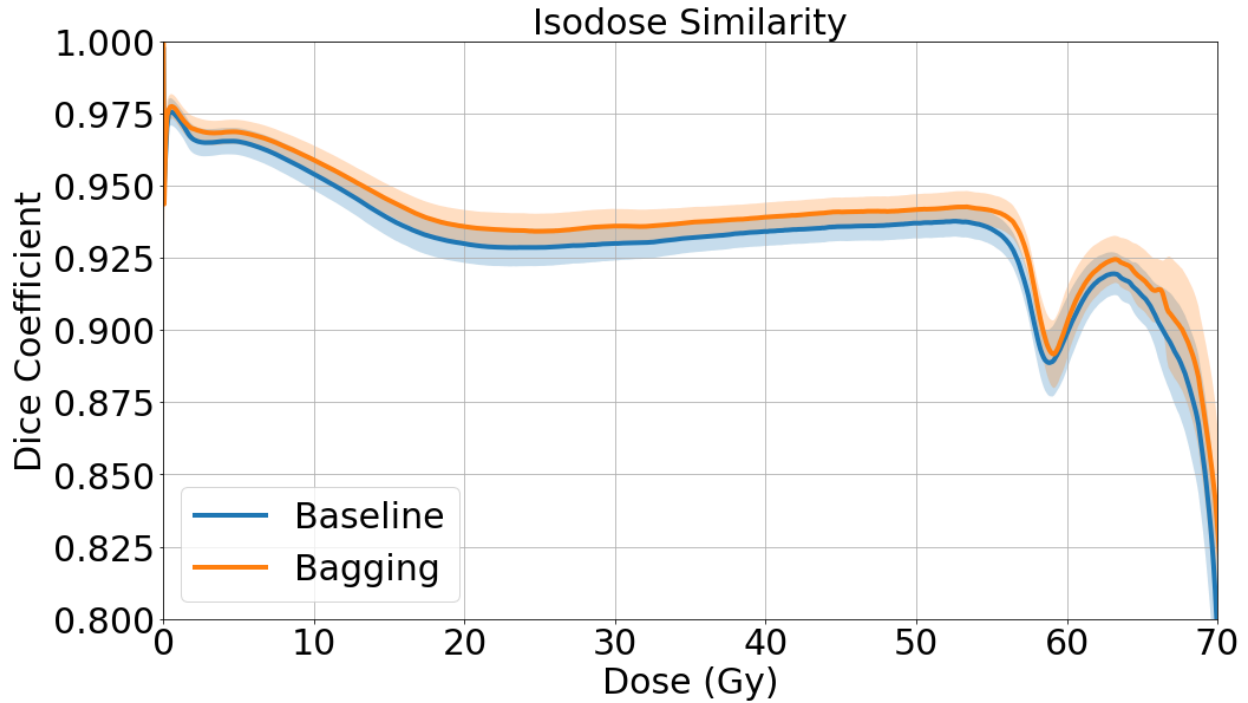

**Figure 6: Isodose similarity between the predictions and ground truth dose. This is found by calculating the Dice coefficient between the $x\%$ isodose contours of predicted and ground truth doses, and is repeated for where $x$ ranges from 0 to 100 percent of the prescription dose of 70 Gy. The solid line represents the mean value found from all 100 test patients and the error band represents the 95% confidence interval $\left(\bar{x} \pm 1.96 \frac{\sigma}{\sqrt{n}}\right)$.**

Figure 6 illustrates the isodose similarity between the predictions and ground truth dose. Averaged over all the patients and isodose levels, the overall Dice coefficient was found to be 0.932 and 0.938 for the baseline and bagging model's respectively. The improvement in the isodose similarity for the bagging framework was found to be statistically significant (p < 0.05). We found that both methods exhibited similar patterns in the isodose similarity curve, such as a large dip between 56 Gy and 63 Gy, likely due to high dose gradients and irregular shaped PTV56 and PTV63 contours.

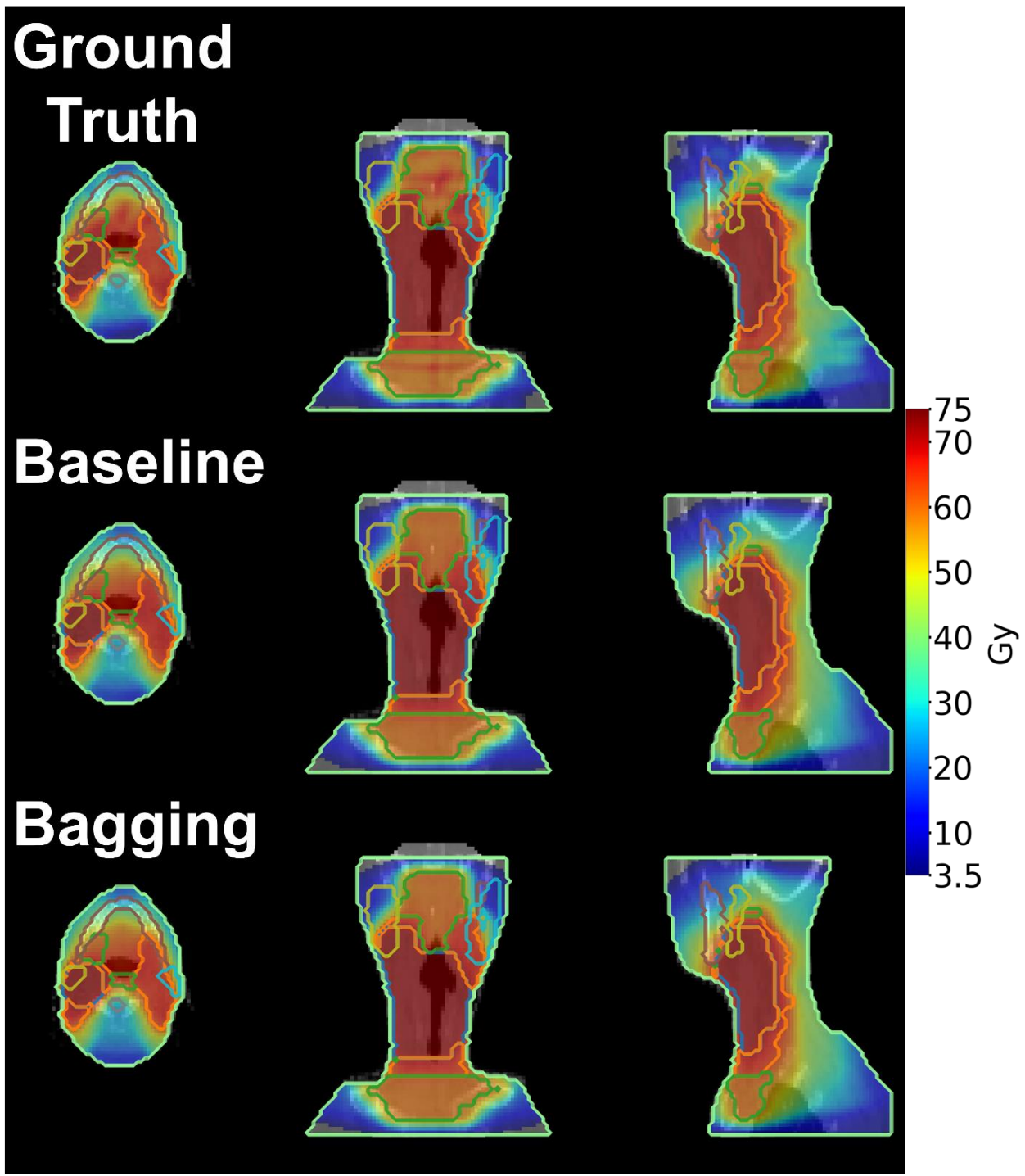

Figure 7: Dose washes of an example test patient for the ground truth (top), baseline prediction (middle), and bagging prediction (bottom). Dose viewing cutoff was set to 3.5 Gy, which is 5% of the prescription dose.

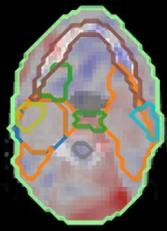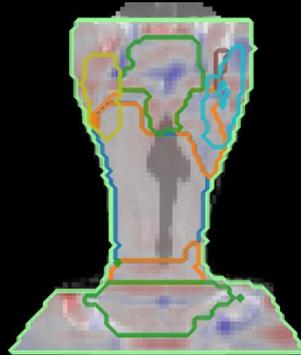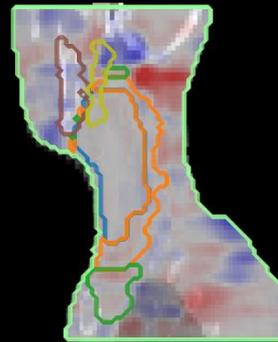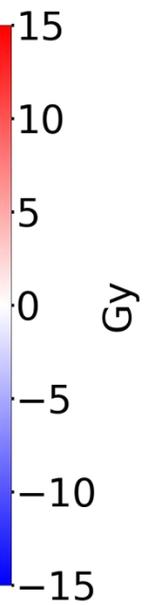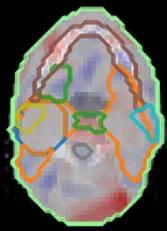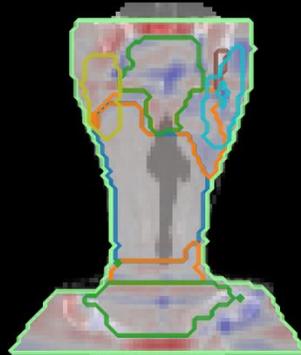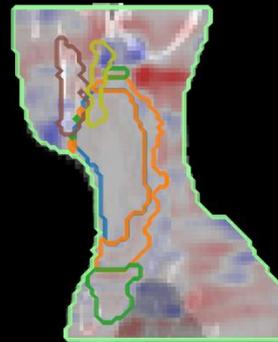

**Figure 8: Voxel-wise dose errors (Prediction – Ground Truth) for the example test patient. Red represents that the model is overpredicting and blue represents that the model is underpredicting.**

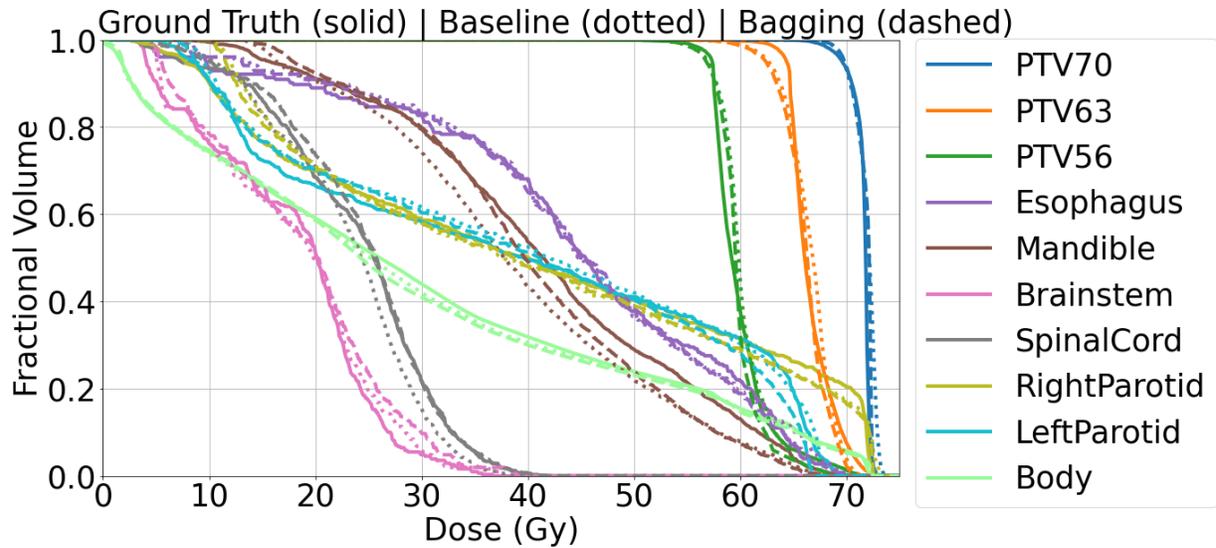

**Figure 9: Dose volume histogram of the ground truth (solid), baseline prediction (dotted), and bagging prediction (dashed) for the example test patient.**

Figure 7 shows the dose washes for the ground truth, baseline prediction, and bagging prediction for a test patient. Visually, all 3 looked similar, but with the predictions losing some low dose detail on the posterior side of the patient. Figure 8 shows the voxel-wise error (Prediction – Ground Truth) for the same example test patient. The errors were in similar positions, but with the baseline model having more areas of underpredicting than the bagging framework. This underprediction can be seen in the DVH shown in Figure 9, particularly in the mandible and the spinal cord. There was some heavier overprediction with the bagging framework over the baseline model in the brainstem for this particular patient. The DVHs were visually similar for the other structures for this patient.

## III.B. Uncertainty Estimation

Using 24 Monte Carlo estimations for the baseline, 24 model predictions for the bagging, both methods were able to create an uncertainty estimation in about the same time frame, which was less than 12 seconds since one evaluation took just under half a second to complete for either framework. Model loading times on and off the GPU was insignificant, since model weights are only 30 MB.

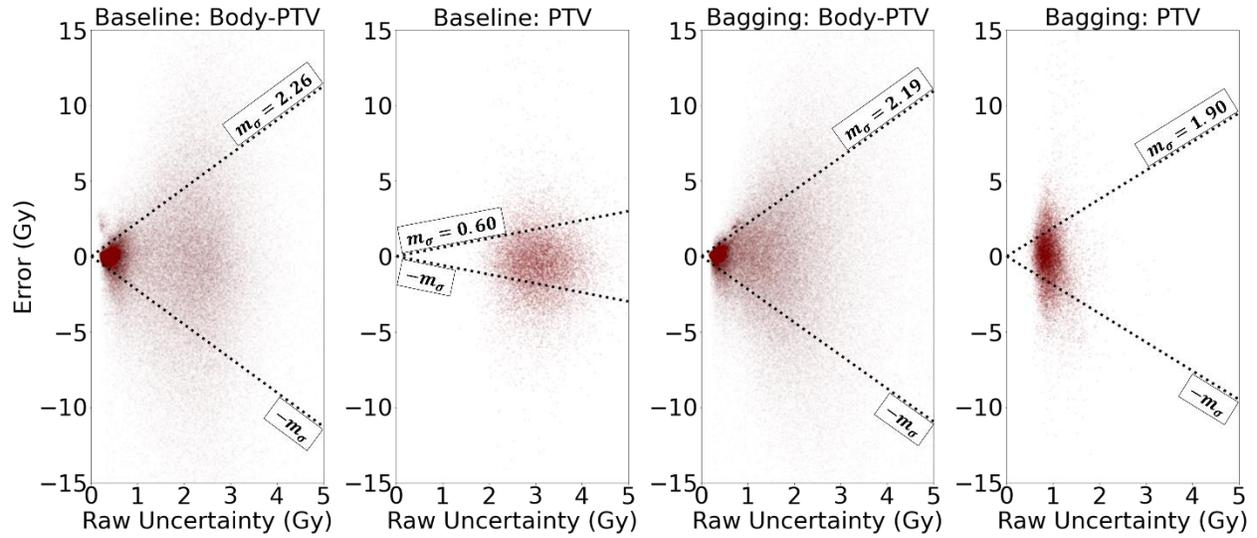

**Figure 10: Error versus uncertainty scatter plot. The slopes of the upper and lower black dotted line ($y = mx$) is $m_\sigma$ and $-m_\sigma$, respectively, which was calculated from the validation data using Equations 3 and 4. The scatter plot illustrates all of the voxels in the test dataset, which is 12,458,654 voxels for the (Body – PTV) and 1,573,697 for the PTV. A transparency alpha value is applied so that it is possible to visualize the density of points.**

Figure 10 shows a scatter plot of the absolute error vs the raw uncertainty estimation. From applying Equations 3 and 4 on the validation data, we found the scaling factors to be $m_{\sigma,(Body-PTV),Baseline} = 2.26$, $m_{\sigma,PTV,Baseline} = 0.60$, $m_{\sigma,(Body-PTV),Bagging} = 2.19$, $m_{\sigma,PTV,Bagging} = 1.90$. By scaling the uncertainty with $m_\sigma$, we find that the voxel-wise scaled uncertainty is greater than the absolute error for 70.8% (baseline) and 71.3% (bagging) of the voxels in the test patients. These values were similar to that of the validation patients, which was 71.6% (baseline) and 71.8% (bagging), indicating that $uncertainty_{scaled} = m_\sigma * uncertainty_{raw}$ does closely represent the same percentage of data.

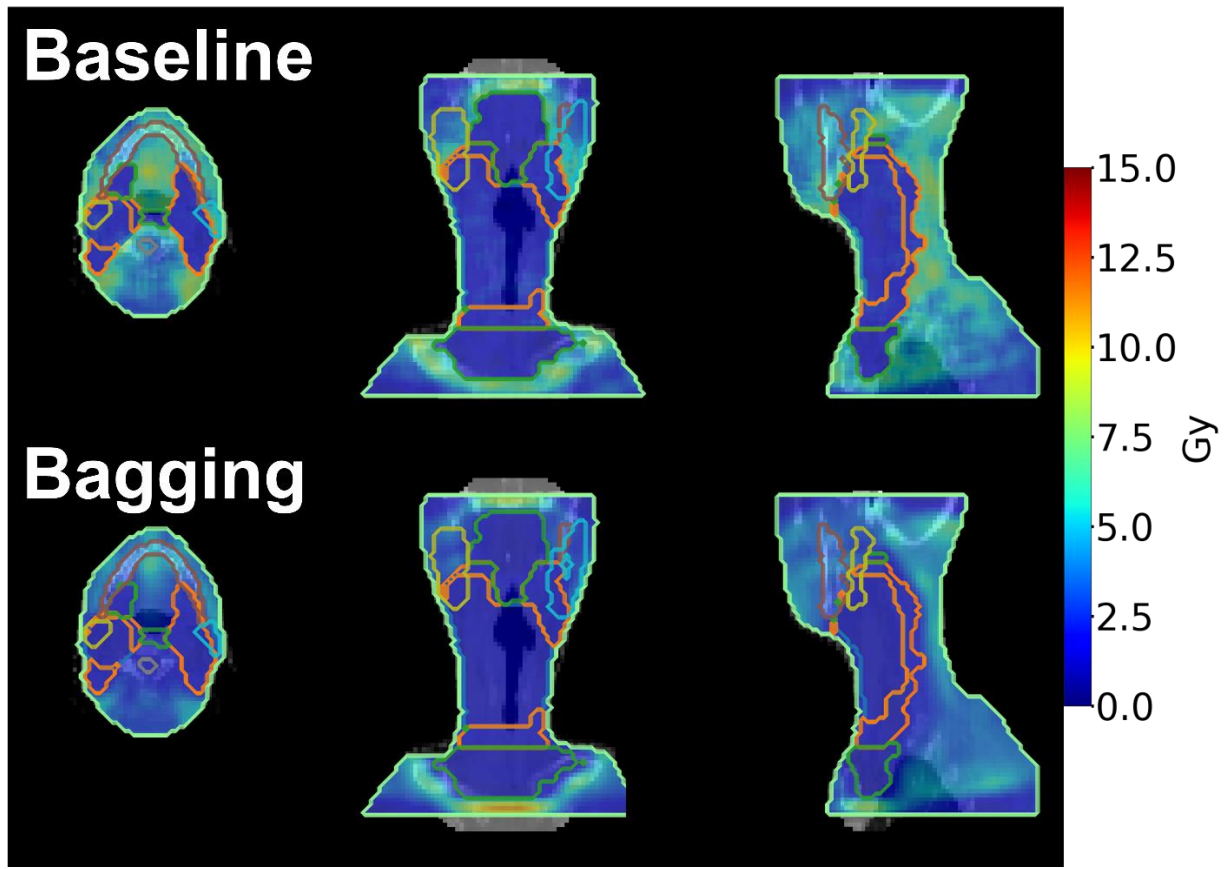

**Figure 11:** Wash of the scaled uncertainty ($m_\sigma * uncertainty_{raw}$) for the example test patient.

From Figure 11, we can see that there are certain similarities in the scaled uncertainties provided by the baseline model and bagging framework, such as low uncertainty in the PTV regions and higher uncertainty in certain areas such as the posterior neck in the sagittal slice. As a whole, the baseline model provided overall larger values of uncertainty than the bagging framework. The Pearson correlation coefficient between the absolute error and the uncertainty was found to be 0.55 (Body–PTV) and 0.27 (PTV) for the bagging framework, and 0.47 (Body–PTV) and 0.026 (PTV) for the baseline model with MCDO, indicating that the bagging framework had an overall better correlation between its prediction error and uncertainty.

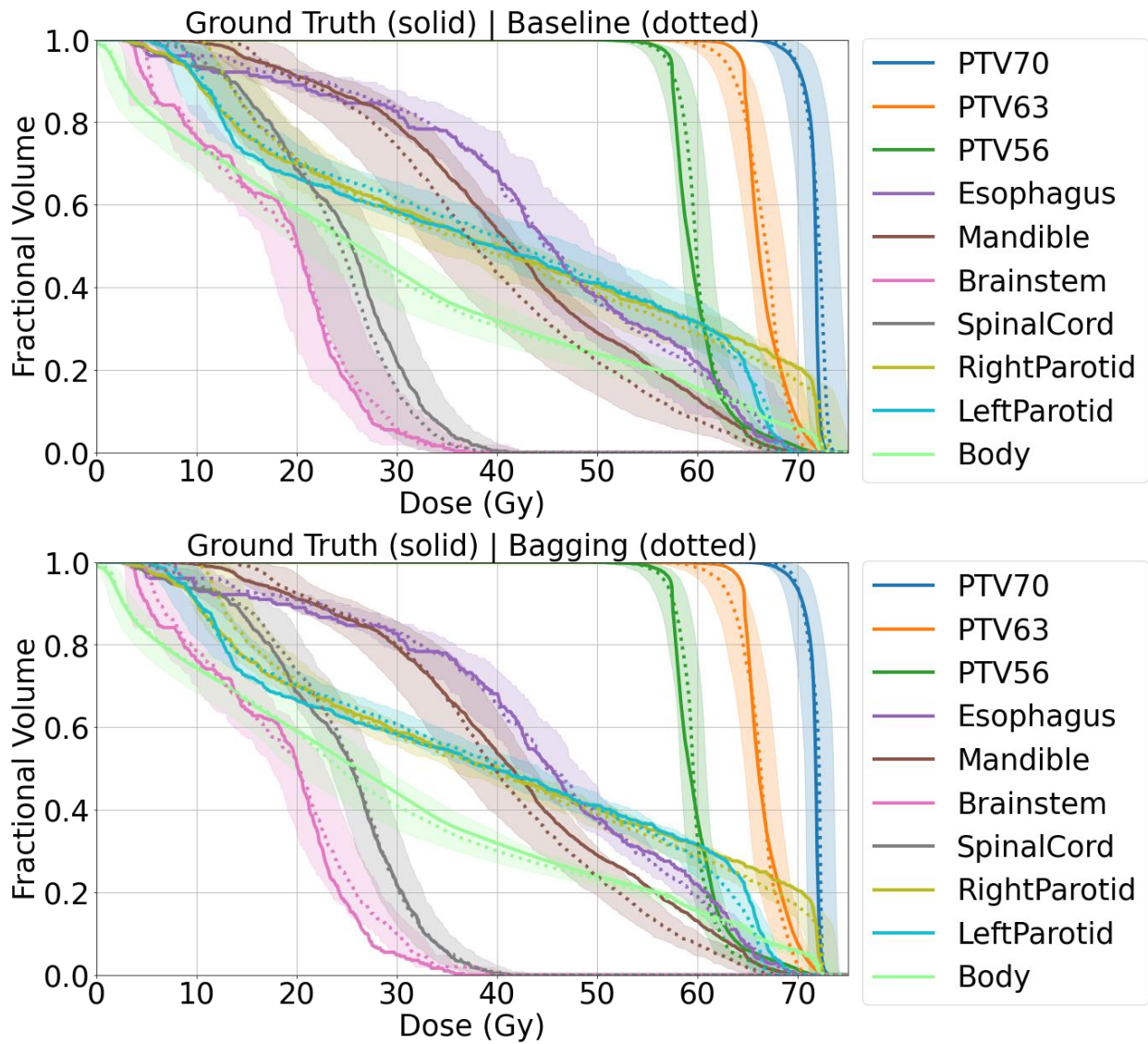

**Figure 12: Dose volume histogram (DVH) plots with uncertainty lower bounds and upper bounds of the example test patient. Bounds are calculated finding the $prediction \pm m_\sigma * uncertainty_{raw}$ in the dose distribution domain and then calculating the DVH.**

Figure 12 shows the DVH plot with additional upper and lower bounds by adding and subtracting the scaled uncertainty, respectively. Overall, the bounds were tighter with the bagging model, which coincides with the overall lower uncertainty values on Figure 12. Since the bounds represent 1 standard deviation, accounting for roughly 70% of the voxel data, there may be some points on the ground truth dose that exceeds the bounds, such as the brainstem on the bagging model.

## IV. Discussion

To our knowledge this is the first DL-based radiotherapy dose prediction study that investigates the use of bagging and uncertainty estimations directly on the volumetric dose distribution. Both the baseline model with MCDO and the bagging framework are capable of producing a reasonable uncertainty estimation. We created a useful method for converting the uncertainty into an absolute metric, so that bounds can be applied with a reasonable interpretation. Overall, with both methods' scaled uncertainties encapsulating the same percentage of voxels, bagging does provide tighter bounds for its scaled uncertainty, as shown in Figure 12. For clinical deployment, we may choose to further adjust the scaling factor, such as setting $m_\sigma \to 2 * m_\sigma$, depending on the conservativeness or aggressiveness of the physician or clinic. Regarding the correlation to uncertainty and absolute error, for both the bagging framework and the baseline model, the area inside the PTV had a lower correlation than outside the PTV. This may suggest that the exact scaling values used may be different for the two regions, in order to get the best performance of the model in a clinical setting.

During the training process, we utilized the Huber loss as our objective function, which is a common surrogate for MAE. The main motivation for Huber loss is that, for the large errors, it acts exactly like MAE, but is additionally differentiable at zero. This differentiability is a desirable trait in optimization for improved stability in convergence without additional heuristics such as learning rate decay. However, we performed much of the performance analysis in MAE, with the error in units of Gy. This allows for us to directly interpret the result in how the model performs clinically in Gy. Huber loss values are less directly interpretable in this sense, since small errors follow a mean squared error instead.

We have shown that, although each individual bagging model sees fewer training patients (160 patients) than the baseline model (200 patients), the ensembling effect overall significantly improves the predictive performance for the bagging model. The only metrics that the averaging of bagging caused to have less prediction accuracy are the PTV $D_{max}$, homogeneity, and conformity, which are more sensitive to the hot and cold spots in the PTV. However, the because of bagging's averaging effect, it reduced the $D_{max}$ hot spot in the PTV, and improved conformity and homogeneity. If used as a clinical guidance tool, this may actually be beneficial, since the planner may have to place more effort in creating a plan that meets higher criteria. Bagging does come at a substantially increased initial cost of training every singly bagging model individually. In our case, the total hours it took to train the total bagging framework was almost 1400 hours, or close to 2 months, as opposed to around 58 hours, about 2.5 days, for just the baseline model. However, since evaluation stage is fast, under half a second for a single prediction, the bagging framework can produce an output and its uncertainty in under 12 seconds, which is clinically reasonable. The baseline model, while being able to output its prediction in under 0.5 seconds, still needs the additional 12 seconds to generate an uncertainty map from 24 Monte Carlo estimations. In terms of deployment with uncertainty estimations turned on, both frameworks offer the same performance time.

The OpenKBP dataset uses 9 field IMRT plans, and thus the trained models in this study are directly applicable to this modality. While the MCDO and bagging methods presented in this paper are applicable to other radiotherapy treatment modalities, such as VMAT, the models produced in this study are not. The creation of models that can be applied to other modalities would require retraining on a new cohort of patients that were treated with that specific modality. Transfer learning methods may be used to adapt the models in this study to a new modality, which can help reduce the number of patients needed to train the model.

The format of the data publicly provided by the OpenKBP organizers had a voxel resolution of 128 x 128 x 128, with a mode voxel spacing of 3.906 mm x 3.906 mm x 2.5 mm. This may be considered a coarse representation of the data, considering that typical CTs natively have sub-millimeter pixel spacing on each slice, and patients can be aligned for treatment planning with millimeter-level accuracy. From a deep learning standpoint, 128 x 128 x 128 as a data input size is large, and, depending on the model architecture specifics, may lead to out-of-memory issues on the GPU. It is expected that as the hardware advances, it will be possible to train future models with much finer data representations.

Currently, the deep neural network predicts the 3D dose distribution, which could be used directly as a clinical guidance tool for planning or generation of plan directives. A machine-deliverable plan can be generated in one of two ways: 1) a treatment planner can use the predicted dose and create a plan that mimics the predicted dose in the commercial treatment planning system, or 2) an automatic dose-mimicking optimization method, such as TORA (Long *et al.*, 2018), can be used to generate a plan from the predicted dose.

It should be noted, that as an ensemble-based metaheuristic, bagging does not compete with any of the previously proposed deep learning architectures used in previous literature for dose prediction in radiotherapy. In fact, any of the previously proposed architectures, whether it be U-net, generative adversarial networks (GAN), support vector machine (SVM), or even basic linear regression models, can additionally use bagging for improved performance. In addition, any of the deep learning models that have previously trained with dropout—which, at this point, most of deep learning models should be using—MCDO can be freely applied during the evaluation phase to produce an uncertainty map, without any need to tweak the existing model.

The decision as to whether the additional computational cost of training a bagging framework is worth the performance benefit or not will be up to the individual use case. It highly depends on the problem and the acceptable predictive error for deployment. Due to the diminishing returns effect of adding more models in the bagging framework, the additional performance benefit of bagging can be estimated by training a few bagging models and extrapolating a curve. This can be used to decide the total number of bagging models necessary for the problem. For our problem in radiotherapy dose prediction, we have found that the baseline model does already produce fairly low errors, on average, of less than 3% for $D_{mean}$ and 4% for $D_{max}$ to structures. The addition of bagging was able to further reduce these errors by another 0.34% for $D_{mean}$ and 0.19% for $D_{max}$, on average. We may see further benefits of bagging if we directly use domain specific losses, or on datasets that are less clean and more variable than the OpenKBP dataset. This will be a topic of further investigation in a future study.

## V. Conclusion

In this study, we propose to use Monte Carlo dropout (MCDO) and the bootstrap aggregation (bagging) technique on deep learning models to produce uncertainty estimations for radiation therapy dose prediction. We show that both models are capable of generating a reasonable uncertainty map, and, with our proposed scaling technique, creating interpretable uncertainties and bounds on the prediction and any relevant metrics. Performance-wise, bagging provides statistically significant reduced loss value and errors in most of the metrics investigated in this study. The addition of bagging was able to further reduce errors by another 0.34% for $D_{mean}$ and 0.19% for $D_{max}$, on average, when compared to the baseline model. The usefulness of bagging does highly depend on the problem and the acceptable predictive error, and its high computational cost for training should be factored in to deciding whether it is advantageous to use it. In terms of deployment with uncertainty estimations turned on, both methods offer the same performance time of about 12 seconds. As an ensemble-based metaheuristic, bagging can be used with existing machine learning architectures to improve stability and performance, and MCDO can be applied to any deep learning models that have dropout as part of their architecture. The dataset we used for this study is publically available from the OpenKBP – 2020 AAPM Grand Challenge.

# VI. Acknowledgements

This study was supported by the National Institutes of Health (NIH) R01CA237269 and the Cancer Prevention & Research Institute of Texas (CPRIT) IIRA RP150485.